\documentstyle[prd,aps,floats]{revtex}
\begin{document}
\draft

\input epsf \renewcommand{\topfraction}{0.8} 
\newcommand{\beq}{\begin{equation}}
\newcommand{\eeq}{\end{equation}}
\newcommand{\pbar}{\not{\!\partial}}
\newcommand{\dbar}{\not{\!{\!D}}}
\def\lsim{\:\raisebox{-0.75ex}{$\stackrel{\textstyle<}{\sim}$}\:}
\def\gsim{\:\raisebox{-0.75ex}{$\stackrel{\textstyle>}{\sim}$}\:}
\twocolumn[\hsize\textwidth\columnwidth\hsize\csname 
@twocolumnfalse\endcsname

\title{Q-ball formation in the wake of Hubble-induced radiative corrections}   
\author{$^{\dagger}$Rouzbeh Allahverdi, $^{\sharp}$Anupam Mazumdar
and $^{\sharp}$Altug Ozpineci}     
\address{$^{\dagger}$ Physik Department, TU M\"unchen, James Frank
Strasse, D-85748, Garching, Germany. \\
$^{\sharp}$ The Abdus Salam International Centre for Theoretical 
Physics,~I-34100~Trieste, Italy.}
\maketitle
\begin{abstract}
We discuss some interesting aspects of the $\rm Q$-ball formation during 
the early oscillations of the flat directions. These oscillations 
are triggered by the running  of soft $({\rm mass})^2$ stemming from 
the nonzero energy density of the Universe. However, this is quite different 
from the standard $\rm Q$-ball formation. The running in presence of 
gauge and Yukawa couplings becomes strong if $m_{1/2}/m_0$ is  
sufficiently large. Moreover, the $\rm Q$-balls which  are formed during the 
early oscillations constantly evolve, due to the redshift of the 
Hubble-induced soft mass, until the low-energy supersymmtery breaking 
becomes dominant. For smaller $m_{1/2}/m_0$, $\rm Q$-balls are not 
formed during early oscillations because of the shrinking of the 
instability band due to the Hubble expansion. In this case the 
$\rm Q$-balls are formed only at the weak scale, but typically carry 
smaller charges, as a result of their amplitude redshift. Therefore, 
the Hubble-induced corrections to the flat directions give rise to  
a successful $\rm Q$-ball cosmology.
\end{abstract}

\pacs{\pacs{PACS numbers: 12.60.Jv;11.30.Fs;98.80.Cq \hspace*{1.3cm}
TUM-HEP-454/02, hep-ph/0203062}}

\vskip2pc]

\section{Introduction}

The presence of {\it flat directions}, generally denoted by $\phi$, in
the field space along which the classical potential vanishes is quite 
generic in the minimal supersymmetric standard model (MSSM) \cite{nilles}.
This has interesting cosmological implications. In particular flat 
directions which are built on squarks and sleptons, and carry a nonzero 
$B-L$, can generate the observed baryon asymmetry in the context of 
Affleck-Dine baryogenesis mechanism \cite{ad,drt1,drt2}. As a second 
off-shoot, the formation of $\rm Q$-balls \cite{coleman,cohen} in 
supersymmetric (SUSY) theories can come from the presence of flat directions
carrying a $U(1)$ charge \cite{kusenko1,dvali,kusenko,em1,em2,ejm}. A
$\rm Q$-ball is a non-topological soliton whose stability is confirmed due 
to the presence of a non-vanishing charge $\rm Q$ it carries associated 
with a $U(1)$ symmetry \cite{coleman,cohen}.

It has been shown \cite{kusenko,em1,em2} that homogeneous oscillations 
of flat directions can be fragmented into $\rm Q$-balls if the flat 
direction potential grows more slowly than $|\phi|^2$, or, equivalently 
if $V(\phi)/|\phi|^2$ has a minimum at $|\phi| \neq 0$. The numerical 
simulations also support this idea \cite{kk}. This can be understood 
by noting that flat direction oscillations for such potentials 
behaves as a fluid with a negative pressure. Therefore, spatial 
inhomogeneities around the zero-mode condensate along the flat 
direction, set by inflationary fluctuations \cite{em1}, can 
exponentially grow. When the modes go nonlinear, a lump of $\rm Q$-matter
forms with a physical size set by the wavelength of the fastest
growing mode. The ${\rm Q}$-ball formation is usually studied in this 
context when the low-energy SUSY breaking triggers the flat direction 
oscillations.

However, all the flat directions of the MSSM are subject to
modification at scales well above the low-energy SUSY breaking
scale. In fact, nonrenormalizable superpotential terms induced by the 
new physics remove the flatness for large $|\phi|$ \cite{drt1,drt2}. 
Furthermore, in supergravity (SUGRA) models, SUSY breaking terms by the 
inflation sector generally induces soft mass terms for 
scalar fields \cite{drt1,drt2,sbi} (unless forbidden by some
symmetry). It is also possible to obtain Hubble-induced gaugino soft
mass $m_{1/2}$, and Hubble-induced $A$-term unless prohibited by
an $R$-symmetry. This has a very important consequence for Affleck-Dine 
baryogenesis as we will explain shortly.

Very recently, the running of the flat direction soft masses when the 
Hubble-induced SUSY breaking is dominant, called the Hubble-induced radiative 
corrections, has been studied \cite{adm}. There, the main focus was on
the viability of the Affleck-Dine baryogenesis in the wake of 
Hubble-induced radiative corrections. The main conclusion was that the 
$H_{\rm u}L$ flat direction is the most promising one. A similar case 
study can be made for the $\rm Q$-ball formation in the early Universe, 
which is the topic of this paper. We show that flat directions made up of 
squarks and sleptons may undergo early oscillations, as a result of 
the Hubble-induced radiative corrections. The ${\rm Q}$-balls 
can be formed during these early oscillations if $m_{1/2}/m_0$ 
is sufficiently large. For smaller values of $m_{1/2}/m_0$, ${\rm Q }$-balls 
are formed when the low-energy SUSY breaking becomes dominant. The main
point here is that during early oscillations the soft masses are 
determined by the Hubble parameter. Therefore, the width of the instability
band is redshifted $\sim H$ which is quicker than the redshift of the
unstable modes $\propto a^{-1}$, where $a$ being the scale factor of
the Universe. If perturbations do not have enough time to grow then 
$\rm Q$-balls are not formed. However, the expansion of the Universe 
reduces the amplitude of oscillations by the time the low-energy SUSY 
breaking takes over. This has interesting consequences for the subsequent
formation of $\rm Q$-balls.

We begin by discussing the dynamics of the flat direction. Then
we briefly describe the Hubble-induced radiative corrections to the
scalar potential. In the subsequent section we discuss $\rm Q$-ball 
formation and consequences of a phase of early oscillations. Finally, 
we conclude our paper.

\section{The flat direction dynamics}

In the early Universe the energy density stored in the inflaton field
\cite{infl} is the dominant
source of SUSY breaking, and induces a $({\rm mass})^2 \propto H^2$ 
for all the MSSM flat directions, where $H$ is the expansion rate of the 
Universe \cite{drt1,drt2,sbi}. The effect of such a mass term crucially 
depends on the size and the sign of the constant of proportionality. 
For a positive  $({\rm mass})^2 \ll H^2$, SUSY breaking by the inflation 
sector has no significant consequences. On the other hand, if 
$({\rm mass})^2 \gg H^2$, the flat direction is heavy enough to
settle down at the bottom of the potential during inflation. 
A stable scalar field might even act as cold dark matter \cite{lm}.

Perhaps, the most interesting case occurs for a $({\rm mass})^2 \sim -H^2$,
since it naturally leads to a nonzero vacuum expectation value (VEV)
for the flat direction at the onset of its oscillations. 
This can be realized at the tree-level in simple extensions of minimal 
supergravity models \cite{drt1,drt2}, and from one-loop corrections to 
the K\"ahler potential in no-scale supergravity models \cite{gmo}.
A detailed examination of the scenario with $({\rm mass}) ^2 \sim -H^2$, 
including a systematic treatment of nonrenormalizable
superpotential terms which lift the flat direction has been performed
in Ref.~\cite{drt2}. The SUSY breaking by the inflaton energy 
density and by the hidden sector result in terms
\begin{eqnarray} 
\label{scalpot}
&-&C_I H^2 {| \phi |}^2 + \left(a {\lambda}_n H {{\phi}^n \over n
M^{n-3}} + {\rm h.c.}\right) + m^{2}_{\phi,0} {| \phi |}^2 \,
\nonumber \\
&&+ \left(A_{\phi,0} {\lambda}_n {{\phi}^n \over n M^{n-3}} + {\rm h.c.}
\right) \,
\end{eqnarray}
in the scalar potential\footnote{For our purpose we consider 
nonrenormalizable superpotential terms $\lambda_{n}\Phi^{n}/nM^{n-3}$, 
where $\Phi$ is the superfield comprising $\phi$ and its fermionic
partner, and $M$ is the scale of new physics which induces such a term. All of
the MSSM flat directions are lifted at $n \leq 9$ level, if there is no
other symmetry except the standard model gauge group
\cite{gkm}.}. The first and the third terms are the
Hubble-induced and low-energy soft mass terms respectively, while the
second and the fourth terms are the Hubble-induced and the low-energy $A$
terms respectively. The Hubble-induced soft terms typically dominate
the low-energy ones for $H > m_{3/2}$, with $m_{3/2}$ being the
gravitino mass. If $C_I > 0$, the absolute value of the flat direction 
during inflation settles at the minimum given by
\begin{equation}
\label{veveq}
|\phi| \simeq \left({C_I \over (n-1) {\lambda}_n} H_I
M^{n-3}\right)^{1/ n-2} \,, 
\end{equation}
with $H_I$ being the Hubble constant during the inflationary
epoch. After the end of inflation, $\langle \phi \rangle$ initially 
continues to track the instantaneous local minimum of the scalar
potential, which can be estimated by replacing $H_I$ by $H(t)$ in
Eq.~(\ref{veveq}). Once $H \simeq m_{3/2}$, the low-energy soft SUSY 
breaking terms take over and $\phi$ starts oscillating. It has 
recently been noticed that various thermal effects can trigger 
oscillations at $H \gg m_{3/2}$ \cite{ace,and}. Here we show that it 
is also possible to obtain a phase of early oscillations from the 
Hubble-induced radiative corrections to the scalar potential.


\section{Hubble-induced radiative corrections}

It has been shown in a recent study \cite{adm}, that the Hubble-induced
radiative corrections can significantly change the shape of the flat
direction potential. This is especially important during and right 
after the end of inflation if the inflationary scale is above the 
weak scale, and if thermal equilibrium is achieved at sufficiently 
late times. Otherwise, thermal corrections to the scalar masses 
\cite{ace,and} always dominate the radiative corrections. We therefore 
consider models with a low reheat temperature which can be naturally
realized in models where the inflation sector is gravitationally
coupled to the matter sector.

All fields which have gauge or Yukawa couplings to the flat direction
contribute to the logarithmic running of its $({\rm mass})^2$.
Therefore, one should study the evolution of the flat direction 
$({\rm mass})^2$ from some higher scale such as $M_{\rm GUT}$ down to low
scales in order to determine the location of the true minimum of the
potential. The running of low-energy soft breaking masses has been 
studied in great detail in the context of MSSM phenomenology \cite{dm}. 
The one-loop beta functions for the $({\rm mass})^2$ of the MSSM scalars 
receive opposite contributions from the scalar and the gaugino loops 
\cite{nilles}. If the top Yukawa coupling is the only large one 
(i.e. as long as $\tan\beta$ is not very large), the beta functions for 
the $({\rm mass})^2$ of squarks of the first and the second generations, 
then the sleptons only receive significant contributions from the 
gaugino loops. A review of these effects can be found in Ref.~\cite{dm}. 
Here we only mention the main results for the universal boundary conditions, 
where at $M_{\rm GUT} \approx 2\times 10^{16}$~GeV the soft $({\rm mass})^2$ 
of all the scalars is $m^2_0$, while the gauginos have a common soft 
mass $m_{1/2}$.

For the first and second generations of squarks 
\begin{equation} 
\label{low1}
m^2 \simeq m^{2}_{0} + (5-7)~m^{2}_{1/2}\,,
\end{equation} 
at the weak scale, while for the right-handed and the left-handed sleptons
\begin{equation}
\label{low2} 
m^2 \simeq m^{2}_{0} + 0.1~m^{2}_{1/2}\, \quad , \quad \quad
m^2 \simeq m^{2}_{0} + 0.5~m^{2}_{1/2}\,,
\end{equation}
respectively. These results are independent of $\tan \beta$, while the
soft breaking $({\rm mass})^2$ of the third generation of squarks has 
some dependence on $\tan \beta$. On the other hand the $({\rm mass})^2$ 
of $H_u$ becomes negative at low scales, e.g.
\begin{equation}
\label{low3} 
m^2\simeq - \frac{1}{2}~m^{2}_{0} - 2~m^{2}_{1/2}\,, 
\end{equation}
for the choice of $\tan \beta = 1.65$ \cite{dm}. The important point is 
that the sum $m^{2}_{H_u} + m^2_L$, which describes the mass of the 
$H_u L$ flat direction, is driven to negative values at the weak 
scale only for $m_{1/2} \gsim m_0$.

Similarly, one could follow the evolution of the soft breaking terms
when the Hubble-induced supersymmetry breaking terms are
dominant\footnote{Apart from the gauge and the Yukawa interactions of the
flat directions there also exist nonrenormalizable interactions coming
from nonminimal kinetic terms in the K\"ahler potential and nonminimal
coupling of the flat directions to the gravity. The flat direction 
$({\rm mass})^2$ receives radiative corrections from all these
interactions. However, at scales well below the Planck scale 
$2\times10^{18}$~GeV, the running from nonrenormalizable interactions 
are taken over by that from the gauge and the Yukawa interactions.
We therefore assume that well below the GUT scale 
it is sufficient to consider the running due to gauge and Yukawa
couplings, while including the running from nonrenormalizable
interactions between $M_{\rm Planck}$ and $M_{\rm GUT}$ in the boundary
condition for $C_I$ at $M_{\rm GUT}$. Roughly speaking, this is equivalent 
to integrating out the heavier modes above the GUT scale.}. However,
the boundary conditions are quite different here, since $m^2_0$ and
$m_{1/2}$ are determined by the scale of inflation, and the form of 
the K\"ahler potential. They become completely negligible at low scales,
and, they have no bearing on present phenomenology, e.g. at present 
the Hubble parameter is $H(0) \sim {\cal O}(10^{-33})$ eV. For more 
details in this regard we refer the reader to Ref.\cite {adm}. Here we 
draw the main conclusions which play important roles in our subsequent 
discussion on the formation of ${\rm Q}$-balls.

A typical flat direction $\phi$ is a linear combination 
$\phi = \sum_{i=1}^N a_i \varphi_i$ of the MSSM scalars $\varphi_i$, 
implying that $m^{2}_{\phi} = \sum_{i=1}^N |a_i|^2 m^{2}_{\varphi}$. As 
mentioned before, for a given $m^2_0$, the running of $m^{2}_{\phi}$ 
crucially depends on $m_{1/2}$. A Hubble-induced gaugino mass can be 
produced from a non-minimal dependence of the gauge superfield kinetic 
terms on the inflaton field.  Generically, the gauge 
superfield kinetic terms must depend on the field(s) of the hidden, or, 
secluded sector in order to obtain gaugino masses of roughly the 
same order, or larger than the scalar masses, as required by phenomenology. 
Having $m_{1/2} \sim H$, thus appears to be quite natural unless an
$R$-symmetry forbids terms which are linear in the inflaton superfield
\cite{drt2}. The same also holds for the Hubble-induced $A$ terms. The
$\mu$ term is a bit different. Since, it doesn't break supersymmetry,
there is a priori no reason to assume that $\mu$ of order $H$ will
be created. As noticed in Ref.\cite{adm}, the viability of
Affleck-Dine baryogenesis only requires $\mu (M_{\rm GUT}) \lsim H/4$.

For $C_{I}\approx -1$, only the $H_{u}L$ flat direction acquires a
negative $({\rm mass})^2$ at low scales. The flipping in the sign 
occurs at a scale $q_{c}$, which is above $\sim {\cal O}(1)$~TeV, 
unless $m_{1/2} < H$. However, the exact value of $q_c$ also depends 
on a number of other model parameters, e.g. $\tan \beta$.

For $C_{I}\approx +1$, the $H_u L$ flat direction always has a 
negative $({\rm mass})^2$ at small scales, but for $m_{1/2} \gsim 3H$ 
the $(\rm mass)^2$ flips its sign twice. The slepton masses only receive 
positive contributions from the electroweak gaugino loops. Therefore, 
the $LLE$ flat direction $({\rm mass})^2$ remains negative down to small 
scales unless $m_{1/2} \gsim 2 H$, in particular $q_c \gsim 10^9$ GeV for 
$m_{1/2} \gsim 3H$. The squared masses of all the squarks (but the
right-handed stop), and with a fair approximation the $U_3 D_i D_j$
and the $LQD$ flat directions, change sign unless $m_{1/2} \lsim H/3$. In
particular, $q_c \simeq 10^{10}~(10^{15})$ GeV, for 
$m_{1/2}/H = 1~(3)$. This is due to the large positive contribution 
from gluino loops to the squared squark masses below $M_{\rm GUT}$. The exact
value of $q_c$ is almost independent of other model parameters in this
case, unlike the $H_u L$ case for $C_I \approx -1$. This is simply
because the top Yukawa coupling has almost no effect on the flat
directions other than $H_u L$. This also suggests that the same
conclusions essentially hold for flat directions built on sleptons and
particularly squarks when $C_I \approx -1$, which we numerically
verified.

The $({\rm mass})^2$ of the flat directions made up of squarks and
sleptons increases towards the lower scales and can be written as
\begin{eqnarray}
\label{mass1}
m^2 &\sim& m^2_0 \left (1+K~{\rm ln} \left 
(\frac{|\phi|^2}{M_{\rm GUT}^2}\right )\right )\,, \quad |\phi| \leq
M_{\rm GUT}\,, 
\end{eqnarray}
where $K$ is a negative constant approximately given by \cite{em1,em2} 
\begin{equation}
\label{gamma}
K \approx - \frac{\alpha}{8\pi}\frac{m^{2}_{1/2}}{m^2_0}\,.
\end{equation}
Here $\alpha$ represents the gauge fine structure constant at the GUT
scale, and $m_{1/2}$, $m^2_0$ are the universal gaugino soft mass and
scalar soft $({\rm mass})^2$ at the GUT scale, respectively. When
the low-energy SUSY breaking is dominant one typically finds, by taking 
phenomenological constraints at the weak scale into account, $|K|
\simeq 0.01-0.1$ \cite{em1,em2}. However, our case at hand is different. 
In the early Universe the Hubble-induced SUSY breaking determines the ratio
$m^2_{1/2}/m^2_0$, which can be $\gg 1$ without affecting the low energy
phenomenology. This is an important point, to which we shall come back
shortly.


\section{Early ${\rm Q}$-ball formation}

\subsection{Cosmological set-up}

During inflation all scalar fields, including the flat directions,
with a mass $m < H$ have quantum fluctuations \cite{infl}
\beq
\delta \phi \sim \frac{H_{I}}{2 \pi}\,,
\eeq
which becomes classical when a particular mode leaves the
horizon. These fluctuations, upon re-entering the horizon,
act as initial seed perturbations which will get amplified when the
flat direction starts oscillating after the end of inflation \cite{em1}.
Note, for our purpose we strictly assume that 
the Universe remains matter-dominated for a sufficiently long time
after the end of inflation. This will be the case for models with a 
low reheat temperature.


\subsection{The flat direction oscillations}

As mentioned earlier, the flat directions built on squarks and/or sleptons
receive a positive contribution from the gaugino loops which increases
their $({\rm mass})^2$ towards lower scales.
For $|m^2_0| \approx H^2$, the rate at which $m^2_{\phi}$ increases 
only depends on $m_{1/2}/H$. For a large enough $m_{1/2}/H$ the flat 
direction $({\rm mass})^2$ exceeds $H^2$ at a large scale $\sim q_c$, 
irrespective of the sign of $C_I$ in Eq.~(\ref{scalpot}), leading to 
early oscillations of the flat direction. In order to examine such a 
possibility we should compare $q_c$ with the instantaneous VEV of the 
flat direction $(H M^{n-3})^{1/n-2}$, see Eq.~(\ref{veveq}).
We note that as long as $q_{c}<(H M^{n-3})^{1/n-2}$, the flat direction 
tracks the instantaneous VEV. Also note that the value
of $q_c$ is fixed by the ratio $m_{1/2}/H$ while the Hubble parameter 
is constantly decreasing in time. Therefore, an overlap eventually 
occurs when
\begin{equation}
\label{overlap}
H_{O} \sim \frac{q_{c}^{n-2}}{M^{n-3}}\,.
\end{equation}
If $m_{3/2} < H_{O} < H_I$, then the flat direction starts 
early oscillations with an initial amplitude $\sim q_c$. Note, that any
overlap during inflation has no interesting consequences, 
because oscillations exponentially die down during inflation. 
A phase of early oscillations is ensured in gravity-mediated SUSY
breaking models if $H_{O} > 10^3$~GeV, while for gauge-mediated models
$H_{O} > 10^{-4}$~GeV can be sufficient, since the mass of the gravitino 
is generically much smaller in these models.

In order to study the possibility of ${\rm Q}$-ball formation during
early oscillations, we briefly recall the analysis of the amplification of
the instabilities \cite{kusenko,em1,em2}. Let us consider an
oscillating homogeneous background
$\phi =(Re^{i \theta})/\sqrt{2}$ with a perturbation 
$\delta R \propto e^{(\alpha(t)+ikx)}$, and 
$\delta \theta \propto e^{(\alpha(t)+ikx)}$. The equations of motion
for the perturbations read \cite{kusenko,em1,em2} 
\begin{eqnarray}
\left(\ddot \alpha +\dot \alpha^2+3H \dot \alpha +\frac{k^2}{a^2}+
V^{\prime \prime}-\dot\theta^2\right)\times \left(\ddot \alpha \,\right. 
\nonumber \\
\left. +\dot \alpha^2+3H \dot \alpha +\frac{k^2}{a^2}+\frac{2\dot R}{R}
\dot \alpha\right)+4\dot \theta^2\dot \alpha^2=0\,,
\end{eqnarray}
where $V$ is the flat direction potential, dot denotes time derivative,
and prime denotes derivative with respect to $R$. If
$\dot\theta^2-V^{\prime \prime}>0$, then there
is an instability band within the momentum range $0<k< k_{\rm max}$, where
\begin{equation}
k_{\rm max} = a \sqrt{\dot \theta^2-V^{\prime \prime}}\,.
\end{equation}
In the above expression $k_{\rm max}$ can either be a constant, or 
increasing. In these cases the modes grow. Otherwise, the
modes are removed constantly from the instability band due to the
momentum redshift.

For quantitative estimates it is sufficient to focus on the case of
gravity-mediated models~\cite{em1,em2}, as this is clearly the situation 
of interest regarding the Hubble-induced SUSY breaking. We first analyze 
the situation when $|K|$ in Eq.~(\ref{mass1}) is smaller than one.
We have verified numerically that $|K| < 1$, if $m_{1/2} < 5H$. The 
behavior of the potential, which is governed by the running mass of the 
flat direction, is given by
\begin{eqnarray}
\label{pot}
V \sim H^2 \left (1 + K~{\rm ln}\left (\frac{|\phi|^2}{q^2_c}
\right )\right) |\phi|^2 \,, \quad |\phi| \leq q_c \,,
\end{eqnarray}      
where we have taken into account the Hubble-induced corrections to
the flat direction. The flat direction $\phi$ starts oscillating about
the origin, with an amplitude $\sim q_{c}$, when $H \sim H_{O}$. Upon 
averaging over oscillations, one obtains an equation of state with a 
negative pressure
\begin{equation}
p \approx \frac{K}{2} \rho \,,
\end{equation}
which results in $k_{\rm max} \approx 2 a~|K|^{1/2} H$ \cite{em1,em2}. 
It takes a finite amount of time for a certain mode to become nonlinear. 
A fairly good estimate of this time scale has been given in 
Refs.~\cite{em1,em2}: 
\beq
\label{time}
\Delta t \approx \frac{10}{|K|}H^{-1}\,.
\eeq      
This is also a good estimate of the time scale for forming a $Q$-ball
from the oscillations. For $|K| < 1$, this time scale exceeds one Hubble time,
therefore several oscillations will be needed to form $\rm Q$-balls. 
However, this is not the end of the story, because
the flat direction $({\rm mass})^2$ is also redshifted. As
a result physical modes within the instability band 
$k_{\rm phys} < |K|^{1/2}H$ are either outside, or at best marginally inside 
the horizon. Furthermore, the width of the instability band 
shrinks more quickly $\propto a^{-3/2}$, compared to the redshift of 
momentum modes $\propto a^{-1}$ in a matter-dominated Universe.
This implies that no subhorizon mode can be made unstable before 
the low-energy SUSY breaking takes over. Therefore, we shall not be
able to form $Q$-balls for $|K| < 1$.

The situation is quite different if $|K| \geq 1$, which is the
case if $m_{1/2} \gsim 5H$. Now, we cannot use the expression in
Eq.(\ref{time}), which was derived when $|K| < 1$, reliably.
On the other hand, if we continue doing so, the time required for
perturbations to become nonlinear is seen to be less than  
the period of oscillations $2 \pi/m$. It may sound  contradictory 
to the whole essence of forming $\rm Q$-balls from negative 
pressure behavior of an oscillating condensate. However, in this 
particular case it is a rapid change in the shape of the potential
which is responsible for the growth of the instabilities. This happens
due to the strong
running of the flat direction $({\rm mass})^2$. As a consequence, the 
fragmentation of the zero-mode condensate occurs very rapidly 
within one oscillation due to nonadiabatic time-variation in 
$m^2_{\phi}$. This we call here an {\it instant} ${\rm Q}$-ball formation. 
This is an interesting possibility, which has not been investigated so far.
This phenomena can not happen when the low-energy SUSY breaking terms are
dominant, because one typically encounters $|K| \lsim 0.1$ \cite{em1,em2,ejm}, 
based on phenomenological constraints at the weak scale.

Finally, we note that for $H > m_{3/2}$, the shape of the potential is
slowly varying in time due to the redshift of $m^2_0$ by the Hubble
expansion. This implies that any ${\rm Q}$-balls 
which are formed during early oscillations must evolve too. It is 
conceivable though, that these ${\rm Q}$-balls are eventually stabilized 
with a size set by the final shape of the potential. Nevertheless, 
their charge and number density can be quite different if they are 
formed at early times. A more precise estimate requires a better 
knowledge of the detailed dynamics of the growth of the perturbations
when $|K|> 1$. This we are lacking at the moment and it goes beyond the
scope of the present paper.

\section{Late-time formation of small $Q$-balls}

In this section, we study two consequences when $|K|<1$, occurring for
$3H \leq m_{1/2} \leq 5H$. As mentioned earlier, this can lead to 
early oscillations, albeit no early formation of $\rm Q$-balls.

\subsection{Baryogenesis}

The scale $q_c$ at which the flat direction $({\rm mass})^2$ becomes 
$\approx H^2$ is mainly determined by the ratio $m_{1/2}/H$, note 
that the dependence is practically exponential. For example, for 
$m_{1/2} \approx 4H$, we find $q_c \geq 3.10^{15}$~GeV for all squarks 
and sleptons, except the right-handed stop. With such a value for 
$m_{1/2}$, for a typical $A$-term; we find $A_{q_c} \sim {\cal O}(H)$, 
almost independent of its boundary value. This is a consequence of the 
gaugino loop contribution to the $A$-term beta function. Moreover, 
a relative phase between $m_{1/2}$ and $A(M_{\rm GUT})$ implies 
that the phase of the $A$-term also runs along with the VEV. 
This is interesting since one can obtain a phase mismatch 
required for the torque, which induces a spiral motion to the flat
direction and consequently generates a baryon asymmetry. We have 
checked that for $m_{1/2} \gsim 3H$ a good yield of baryon asymmetry with 
$n_B/n_\phi \sim 10^{-2}$ can be comfortably accommodated. Once
the Hubble-induced $A$-term is effectively switched-off, the comoving
baryon asymmetry becomes frozen. Our conclusion is that baryogenesis 
during early oscillations only depends on the Hubble-induced SUSY breaking
parameters, and it is independent of the low-energy soft mass and $A$-term.

\subsection{Damping the amplitude of oscillations}

From the onset of oscillations at $H_{O}$, given by Eq.~(\ref{overlap}),
the flat direction simply oscillates until the low-energy SUSY breaking 
effects take over. The amplitude of the oscillations decreases as 
$\propto t^{-1/2}$, while  the redshift of the flat
direction mass is $\propto t^{-1}$. As we have described no 
${\rm Q}$-balls are formed during this interval for $|K| < 1$.
The situation rapidly changes once the low-energy SUSY breaking terms take 
over, because the width of the instability band becomes constant there on. 
Then there is a chance for the subhorizon modes to grow, and 
eventually collapse into lumps of ${\rm Q}$-balls. However, 
the amplitude of flat direction oscillations at that time 
has decreased to a new value
\beq
\label{veveq1} 
|\phi| \sim q_c \left ({H_0 \over H_{O}}\right )^{1/2} \,,
\eeq
due to the redshift during the matter-dominated era,
where $H_0$ determines the Hubble rate at the time of low 
energy SUSY breaking scale. By replacing $H_{O}$ from 
Eq.~(\ref{overlap}), we obtain
\beq
\label{veveq2}
|\phi| \sim q_c \left ({H_0 M^{n-3} \over q^{n-2}_c}\right)^{2/3}.
\eeq
Now, in order to illustrate the significance of dampening of the
amplitude, let us compare it with the amplitude when the oscillations 
had started at $H \simeq H_0$. In the latter case the flat direction 
would have tracked the instantaneous VEV, which at $H \simeq H_0$ is given by
$\phi_{0}\sim (H_0 M^{n-3})^{1/(n-2)}$. The ratio of the two
amplitudes is then given by
\begin{equation}
\label{vevratio}
{|\phi| \over |\phi_0|} \sim \left
[ \frac{(H_0 M^{n-3})^{1/(n-2)}}{q_{c}}\right]^{(n-4)/2}\,.
\end{equation}
Note, that this is less than one if $q_{c} > (H_0 M^{n-3})^{1/(n-2)}$. 
Therefore, early oscillations in general damp the amplitude of the 
oscillations at $H_0$, leading to the formation of smaller ${\rm Q}$-balls.

This has important consequences for ${\rm Q}$-ball cosmology.
Recent detailed analysis by lattice simulations show that almost all 
of the generated baryon/lepton asymmetry is absorbed into the 
${\rm Q}$-balls \cite{kk}. This can be problematic if large ${\rm Q}$-balls 
are formed. In gravity-mediated SUSY breaking models the late decay of
large $\rm B$-balls
below the freeze-out temperature for the lightest supersymmetric 
particle (LSP) annihilation, while generating 
$n_B/n_\gamma \sim {\cal O}(10^{-10})$, results in LSP overproduction 
\footnote{This problem can be alleviated for the wino and Higgsino
dark matter \cite{fh}.} \cite{em2}. Note, that the charge of the $\rm
Q$-balls in
gravity-mediated models is proportional to the square of the amplitude;
${\rm Q}\propto |\phi|^2$ \cite{kk}. Therefore, early oscillations of the flat
direction indeed helps reducing the $\rm Q$-ball charge. As an
example, consider a flat direction which is lifted at the $n = 8$
superpotential level with $M = M_{\rm Planck}$. Then it turns out that
$m_{1/2} \approx 4H$ will be sufficient to trigger early oscillations,
resulting in the formation of $\rm Q$-balls which are $6$ orders of
magnitude smaller. The situation is clearly better for smaller $n$
and/or $M = M_{\rm GUT}$, since a smaller $m_{1/2}/H$ will also help
decreasing the amplitude of the oscillations.

In the gauge-mediated models the SUGRA contribution dominates the
flat direction potential at large VEVs, though with a 
smaller gravitino mass $m_{3/2}$. For example, with 
$m_{3/2} \sim 10^{-4}$ GeV, the potential reads
\begin{eqnarray}
V \sim m_{3/2}^2 \left(1+K \ln \left( \frac{|\phi|^2}
{M^2_{\rm GUT}} \right) \right) |\phi|^2\,,
\end{eqnarray}
for $|\phi| > 10^{10}$ GeV. On the other hand, for $|\phi| \leq
10^{10}$ GeV, the gauge-mediated contribution
\begin{equation}
V \sim m^4_{\rm SUSY} \ln \left(\frac{|\phi|^2}{m^2_{\rm SUSY}}\right)\,,
\quad \quad |\phi| \geq m_{\rm SUSY}\,. 
\end{equation}
dominates the potential, with $m_{\rm SUSY} \sim {\cal O}$(TeV)
\cite{dmm}. In this case
${\rm B}$-balls with a charge $B \geq 10^{12}$ act as a stable dark matter 
candidate in the Universe \cite{kusenko}. However, stable $\rm B$-balls 
overclose the Universe in almost all the regions of the parameter space, 
if they ought to yield an acceptable baryon asymmetry \cite{kk2}.

Our earlier estimates can be repeated to show that the
Hubble-induced radiative corrections indeed lead to early oscillations
which eventually decrease the charge of the $\rm Q$-balls. However,
the reduction in the $\rm Q$-ball charge now depends on where $|\phi|$ starts,
see Eq.~(\ref{veveq2}). If $|\phi| \geq 10^{10}$ GeV,
the SUGRA corrections 
dominate and $Q \propto |\phi|^2$, thus our earlier arguments hold
even for the gauge-mediated models.
On the other hand, for $|\phi| < 10^{10}$ GeV, we have 
${\rm Q} \propto |\phi|^4$ \cite{dvali}. Therefore, early oscillations 
reduce the $\rm Q$-ball
charge even more significantly. As an example, consider a flat
direction lifted at the $n=9$ superpotential level with 
$M = M_{\rm Planck}$. Then for $m_{1/2} \approx 4H$, the early oscillations 
result in the formation of gravity-mediated type $\rm Q$-balls which are 
$10^7$ times smaller\footnote{The gravity-mediated type $\rm B$-balls are
stable in gauge-mediated models, since there are no baryons with a mass 
less than $10^{-4}$ GeV \cite{kk3}.}. For smaller $n$, and/or 
$M = M_{\rm GUT}$, we find $|\phi| < 10^{10}$~GeV from Eq.(\ref{veveq2}),
leading to the formation of even smaller $\rm Q$-balls.

We conclude that for $m_{1/2} \gsim 3H$, the flat directions built on
squarks and/or sleptons in general undergo early oscillations. Then the
amplitude of the oscillations is redshifted by the expansion of the
Universe. This results in the formation of smaller $\rm Q$-balls in
both gravity and gauge-mediated models of SUSY breaking. In 
Ref.~\cite{yanagida} a gauged $U(1)_{B-L}$, which is broken at a scale
$v \sim 120^{14}$ GeV, has been invoked to
reduce the charge of a $\rm Q$-ball. In the 
gravity-mediated models the $D$-term from $U(1)_{B-L}$ helps 
forming smaller $\rm B$-balls from the oscillations of the flat directions
with a nonzero $B-L$ at the weak scale. However, as noticed by the
authors, this mechanism
does not improve the situation for the gauge-mediated models. 
The scenario mentioned here works well
for both gravity- and gauge-mediated models, thus
helping the emergence of a successful $\rm Q$-ball cosmology.

\section{Conclusion}

We have considered possible effects of radiative corrections to the
Hubble-induced soft $({\rm mass})^2$ of the flat directions on 
$\rm Q$-ball formation. The key observation is that for a Hubble-induced
soft gaugino mass $m_{1/2} \gsim 3H$, the $({\rm mass})^2$ of
all squarks and sleptons exceeds $H^2$ at a scale $q_c \gsim 10^{15}$
GeV, and for $m_{1/2} \gsim 4H$, it turns out that $q_c \sim M_{\rm GUT}$. 
Note, that the ratio $m_{1/2}/H$ is not constrained by the weak scale 
physics since the Hubble-induced SUSY breaking has no bearing on present 
phenomenology. Our conclusion is that all MSSM flat directions
start oscillating shortly after the end of inflation, independently
from the mechanism of low-energy SUSY breaking mediation. The
formation of $\rm Q$-balls during early oscillations is different from 
the standard case. For $m_{1/2} > 5H$, the logarithmic running of the 
flat direction $({\rm mass})^2$ is so strong that its nonadiabatic 
time-variation may result in the fragmentation of the homogeneous 
condensate during just one oscillation, dubbed as {\it instant} 
$\rm Q$-ball formation. However, $\rm Q$-balls which may be formed 
in this fashion constantly evolve. This is because of the change in the 
shape of the potential caused by redshift of the flat direction 
$({\rm mass})^2$. It is conceivable that $\rm Q$-balls are eventually 
stabilized with a size set by the final shape of the potential, though 
with a different charge and number density. A more precise estimate 
requires quantitative analysis of $\rm Q$-ball formation in this regime.

For $3H \lsim m_{1/2} \lsim 5H$,  the early oscillations in general start,
but no $\rm Q$-balls can be formed due to the continuous shrinking of the
instability band caused by the Hubble expansion. The expansion also
redshifts the amplitude of the oscillations leading to the formation
of considerably smaller $\rm Q$-balls at low scales. We noted that for 
$m_{1/2} \gsim 3H$, the generated baryon asymmetry is determined by 
the Hubble-induced $A$ term as it is also running, which can provide 
a phase mismatch required for generating baryogenesis. We
presented examples of the flat directions which are
lifted by nonrenormalizable superpotential terms of higher order and
found a substantial reduction in the charge of the $\rm Q$-balls. 
Therefore, early oscillations can lead to a successful 
$\rm Q$-ball cosmology in both gravity and gauge-mediated SUSY breaking
models.

\section*{acknowledgements}

The authors are thankful to M. Bastero-Gil, D. Chung, M. Drees, 
and K. Narain for helpful discussions on various aspects of this
project. The work of R.A. was supported by
``Sonderforschungsbereich 375 f\"ur Astro-Teilchenphysik'' der
Deutschen Forschungsgemeinschaft.


\end{document}